\documentclass[12pt]{iopart}
  
\def\40K{$^{40}$K}
\def\K{$^{39}$K}
\def\Na{$^{23}$Na}
\def\NaK{\Na\K}
\usepackage{braket}
\usepackage{graphicx}
\usepackage{iopams}
\usepackage{color}

\begin{document}

\title{A pathway to ultracold bosonic \NaK\ ground state molecules}

\author{Kai~K.~Voges, Philipp~Gersema, Torsten~Hartmann,\\ Torben A.~Schulze, Alessandro~Zenesini and Silke~Ospelkaus}

\address{Institut f\"ur Quantenoptik, Leibniz Universit\"at Hannover, 30167~Hannover, Germany}
\ead{silke.ospelkaus@iqo.uni-hannover.de}
\vspace{10pt}
\begin{indented}
\item[]\today
\end{indented}

\begin{abstract}
We spectroscopically investigate a pathway for the conversion of \NaK\ Feshbach molecules into rovibronic ground state molecules via STImulated Raman Adiabatic Passage (STIRAP). Using photoassociation spectroscopy from the diatomic scattering threshold in the $a^3\Sigma^+$ potential, we locate the resonantly mixed electronically excited intermediate states $\ket{B^1\Pi, v=8}$ and $\ket{c^3\Sigma^+, v=30}$ which, due to their singlet-triplet admixture, serve as an ideal bridge between predominantly $a^3\Sigma^+$ Feshbach molecules and pure $X^1\Sigma^+$ ground state molecules. We investigate their hyperfine structure and present a simple model to determine the singlet-triplet coupling of these states. Using Autler-Townes spectroscopy, we locate the rovibronic ground state of the \NaK\ molecule ($\ket{X^1\Sigma^+, v=0, N=0}$) and the second rotationally excited state $N=2$ to unambiguously identify the ground state. We also extract the effective transition dipole moment from the excited to the ground state. Our investigations result in a fully characterized scheme for the creation of ultracold bosonic \NaK\ ground state molecules.

\end{abstract}

\section{Introduction}
\label{sec1}
\begin{figure}[ht]
	\includegraphics*[width=1\columnwidth]{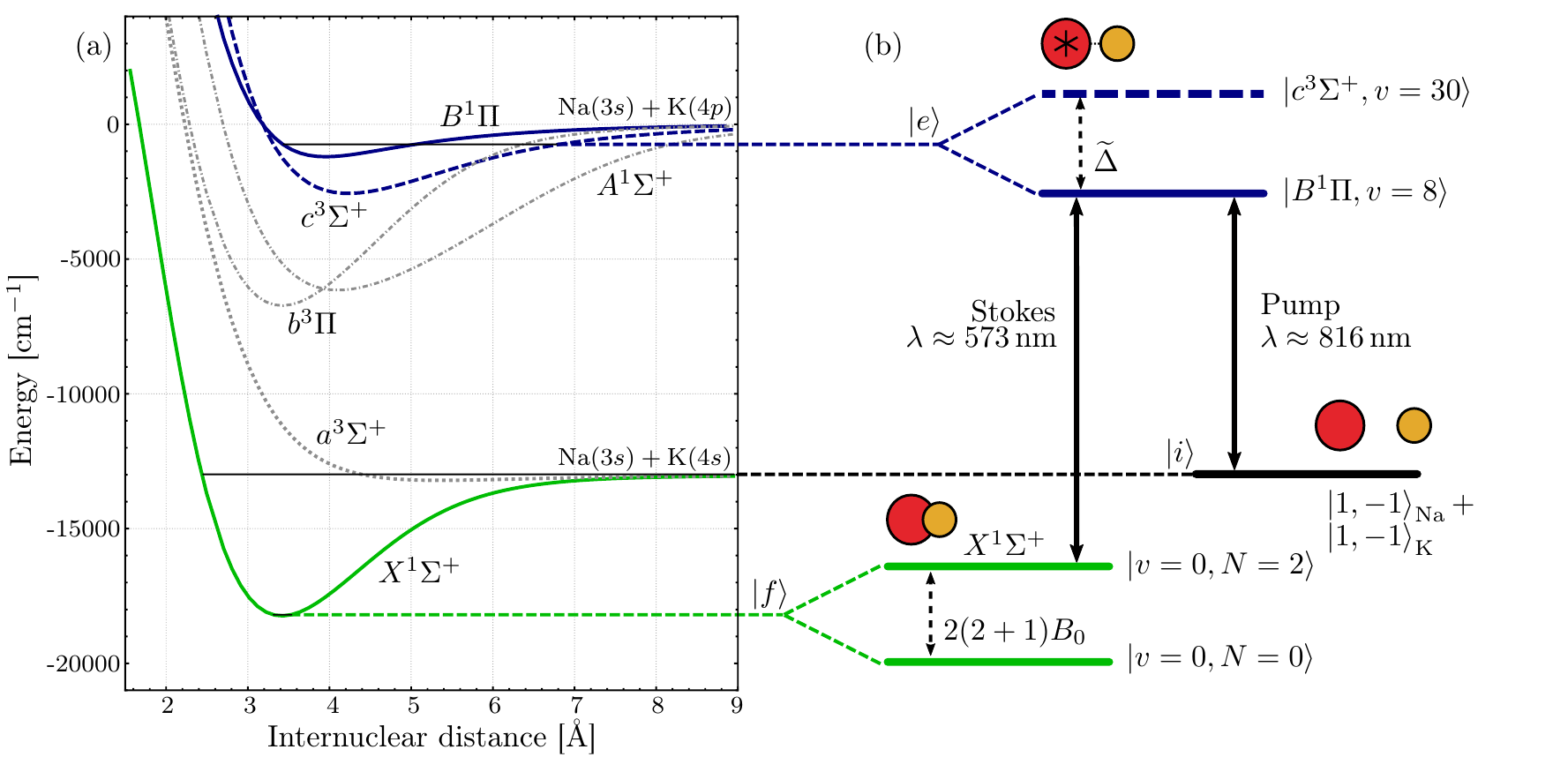}
	\caption{Molecular $\Lambda$-level scheme for the \NaK. (a) The potential energy curves for the NaK molecules which are involved in the spectroscopy (blue and green lines). The gray dashed and dashed dotted lines indicate energetically deeper electronic potentials which can be used for alternative STIRAP pathways (see text). Curves from \cite{Aymar2017}. (b) A simplified sketch displays the energy levels. Red and yellow full circles visualize free \K\ and \Na\ atoms  respectively ($\ket{i}$), the excited molecule ($\ket{e}$) and the ground state molecule ($\ket{f}$).  The black line refers to the atomic state of the \Na+\K\ mixture ($\ket{i}$). The dark blue lines in the upper part are the particular vibrational states in the triplet and singlet state manifold ($\ket{e}$). The energy shift between these states is given by the perturbed energy difference $\widetilde\Delta$ (see Sec.\ref{ExState}). The green lines represent the electronic ground state ($\ket{f}$) with its rotational ground state $N=0$ and the excited state $N=2$ which are separated by $6\times B_0$. The solid arrows indicate the light from the spectroscopy referred to as Pump and Stokes laser light.}
	\label{Potentials}
\end{figure}

Quantum gases of ultracold polar molecules offer unprecedented novel opportunities for the investigation of dipolar collisions, quantum chemical processes and quantum many-body systems \cite{IntroductionKRbChemicalReactionOspelkaus2010, WangCollisions, Inelastic2010,QuantumMagnetPolar,ThreeInteractionPolar}. The new handle in comparison to atomic systems is given by the electric dipole moment of heteronuclear diatomic ground state molecules. 
The most successful approach to the creation of ultracold ensembles of polar ground state molecules is based on the association of two chemically different ultracold atomic alkali species. This process starts with magneto-association to weakly bound Feshbach molecules and continues with the coherent transfer of the weakly bound molecules to the rovibrational ground state of polar molecules using a STIRAP \cite{STIRAPBergmann}.
For the transfer from the initially magneto-associated molecular state $\ket{i}$ to the final molecular ground state $\ket{f}$, the STIRAP transfer consists of an adiabatic change of the dressed state composition by involving two coherent laser beams referred to as Pump and Stokes laser. The Pump laser couples $\ket{i}$ to a third excited state $\ket{e}$, the Stokes laser $\ket{f}$ to $\ket{e}$, resulting in a typical $\Lambda$-level scheme; see Fig.\ref{Potentials}.  During the time evolution the intermediate state $\ket{e}$ is not populated and therefore does not contribute to incoherent molecule losses due to spontaneous decay. In the case of bi-alkali molecules, the coupling between $\ket{i}$/$\ket{f}$ and $\ket{e}$ is governed by the Franck-Condon overlap, the singlet-triplet fraction and the hyperfine composition of the states. For bi-alkali heteronuclear molecules the ground state $\ket{f}$ always belongs to the $X^1\Sigma^+$ potential while a weakly bound dimer state $\ket{i}$ exists mainly in the $a^3\Sigma^+$ potential; see Fig.\ref{Potentials}(a). To act as an efficient coupling bridge, the choice of  $\ket{e}$ is crucial. It needs to be chosen to couple well to the initial triplet as well as to the final singlet state. This can be achieved through the careful choice of mixed states in the $^3\Pi/^1\Sigma^+$ or $^3\Sigma^+/^1\Pi$ potentials; see Fig.\ref{Potentials}(a). Up to now, the production of trapped ground state molecules, either fermionic, such as $^{40}\textrm{K}^{87}\textrm{Rb}$ \cite{KRb1}, $^{23}\textrm{Na}^{40}\textrm{K}$ \cite{GsDiMo23Na40K2015} and $^{6}\textrm{Li}^{23}\textrm{Na}$ \cite{GsDiMo23Na6Li2017}, or bosonic, such as $^{87}\textrm{Rb}^{133}\textrm{Cs}$ \cite{GsDiMo87Rb133Cs2014Grimm,GsDiMo87Rb133Cs2014Cornish} and $^{23}\textrm{Na}^{87}\textrm{Rb}$ \cite{NaRb1} has been reported.  NaRb \cite{NaRb1} and RbCs \cite{GsDiMo87Rb133Cs2014Grimm,GsDiMo87Rb133Cs2014Cornish} ground state molecules have been prepared by using coupled states in the $A^1\Sigma^+$ and $b^3\Pi$ potentials. KRb \cite{KRb1} and fermionic $^{23}\textrm{Na}^{40}\textrm{K}$ \cite{GsDiMo23Na40K2015} ground state molecules have been created using states in the $c^3\Sigma^+$ and $B^1\Pi$ potentials. Additionally, ground state molecules for fermionic $^{23}\textrm{Na}^{40}\textrm{K}$ have been created using excited states in energetically higher $D^1\Pi$ and $d^3\Pi$ potentials \cite{Frauke}; not shown in Fig.\ref{Potentials}(a).\\
In the case of \NaK, suitable transitions for the creation of ground state molecules have so far only been investigated experimentally in hot beam experiments accessing rotational states with $N\ge6$ \cite{NaKSpectroscopyTiemann2015}. Moreover, the work \cite{NaKSpectroscopyTiemann2015} does not include hyperfine structure and offset magnetic fields. In addition, previous theory work \cite{Schulze} describes a complete spectrum involving spin-orbit, Coriolis and spin-rotation interactions as well as Franck-Condon factors identifying promising singlet-triplet mixed states. However, the hyperfine structure still remains untreated although it is crucial for a successful STIRAP transfer \cite{Frauke}. In this paper, we present hyperfine resolved spectroscopic investigations of the \NaK\ molecule at bias magnetic fields for strongly mixed states known from \cite{NaKSpectroscopyTiemann2015,Schulze}.
\\\noindent
Using one-photon association spectroscopy, we first locate and characterize the strongly mixed  $\ket{B^1\Pi, v=8}$ and $\ket{c^3\Sigma^+, v=30}$ states serving as a bridge between the $a^3\Sigma^+$ and $X^1\Sigma^+$ potentials. A large Zeeman splitting at $150\,\textrm{G}$ enables us to perform hyperfine resolved optical spectroscopy of the excited molecular state manifold. We present a simple model to determine the singlet-triplet admixtures of the excited states. Furthermore, we perform two-photon Autler-Townes spectroscopy \cite{AutlerTownes} to locate the rovibronic ground state, determine the rotational constants and extract effective dipole transition matrix elements from our measurements. Thus, we present and fully characterize a pathway for the creation of rovibronic ($v=0, N=0$) \NaK\ ground state molecules for the first time.
\\\noindent
In the following, we give an outline of the experimental setup and procedures; see Sec.\ref{ExpSet}. In Sec.\ref{LocMod} we present the model fit for the coupled excited states, and in Sec.\ref{OnePhot} we detail the spectroscopic one-photon measurements. This is followed by the two-photon ground state spectroscopy which includes the determination of the rotational constant and the transition dipole element of the Stokes transition in Sec.\ref{TwoPhoton}.

\section{Experimental setup and procedures}
\label{ExpSet}
The spectroscopic studies presented in this work are based on the preparation of an ultracold mixture of \Na+\K. A description of the experimental setup can be found in \cite{Gempel,Schulze,Hartmannphd}. Details of the experimental procedure can be found in \cite{SchulzeBEC2018,Hartmann1}. In the following, we briefly summarize the main experimental steps.  First, \Na\ and \K\ atoms are loaded into a two-color magneto-optical trap, followed by  simultaneous molasses cooling of both species. The atoms are optically pumped to the $F=1$ manifold and loaded into an optically plugged magnetic quadrupole trap, where forced microwave evaporative cooling of \Na\ is performed. \K\ atoms are sympathetically cooled by \Na. After both atomic species are transferred into a $1064\,\textrm{nm}$ crossed-beam optical dipole trap (cODT), a homogeneous magnetic field of $150\,\textrm{G}$ is applied to ensure favorable scattering properties in the mixture \cite{SchulzeBEC2018}. The atoms are prepared in the $\ket{f=1, m_f=-1}_\textrm{\scriptsize Na} + \ket{f=1, m_f=-1}_\textrm{\scriptsize K}$ states. Then, forced optical evaporation is performed on both species by lowering the intensity of both beams of the cODT. The forced evaporation process is stopped when the atomic sample has a temperature of about $1\, \mu\textrm{K}$ and a corresponding phase space density $\le\,0.1$ for both species. To decrease the differential gravitational sag and assure a significant overlap of the two clouds, the cODT intensity is ramped up again. For the spectroscopy presented in this work, the magnetic field is either set to $130\,\textrm{G}$ or left at $150\,\textrm{G}$ yielding values at which the inter and intra species scattering properties allow for a long lifetime of the sample by minimizing three-body losses. This allows for a sufficiently long interaction time during photoassociation experiments, as required by the expected weak coupling between the atom pair at the scattering threshold and the electronically excited state molecules.\\\noindent
The laser light for the Pump and the Stokes transitions at $816\,\textrm{nm}$ and $573\,\textrm{nm}$ respectively is generated by diode laser systems. The $816\,\textrm{nm}$ light is generated by a commercial external-cavity diode laser (ECDL). The generation of the $573\,\textrm{nm}$ light starts from a commercial ECDL operating at $1146\,\textrm{nm}$. The light is amplified by a tapered amplifier and subsequently frequency doubled in a self-built resonant bow-tie doubling cavity. Both lasers are locked to a commercial ultra-low expansion (ULE) glass cavity by using the Pound-Drever-Hall technique \cite{Drever1983}. A sideband locking scheme involving widely tunable electro-optical modulators ensure the tunability of both locked lasers within the free spectral range of the ULE cavity which is $1.499\,\textrm{GHz}$. The finesse of the ULE cavity is 24900 and 37400 for $816\,\textrm{nm}$ and $1146\,\textrm{nm}$ light, respectively. The laser system setup is similar to the one described in \cite{Gregory_2015}.\\
Optical fibers spatially filter the light and ensure high quality beam properties. The polarization is set by a half-wave plate for each wavelength. The foci of the two beams on the atoms are adjusted to 1/$e^2$-beam waists of $35\,\mu\textrm{m}$ for the Pump laser and $40\,\mu\textrm{m}$ for the Stokes laser. The maximum laser power is $25\,\textrm{mW}$ for each laser. With the Pump beam at full power, the Stokes beam is reduced to $\le 10\,\textrm{mW}$ to avoid a depletion of \Na\ atoms in the trap center, originating from a strong repulsive dipole force. The beams are geometrically superimposed and orientated perpendicularly to the applied magnetic field.\\\noindent

\section{Excited state spectroscopy}
\label{ExState}
\subsection{Local model for the excited state manifold}
\label{LocMod}
In our experiment, we are specifically interested in a detailed understanding of the previously located \cite{NaKSpectroscopyTiemann2015} strongly mixed $\ket{B^1\Pi, v=8}$ and $\ket{c^3\Sigma^+, v=30}$ states. Hyperfine splitting and spin-orbit coupling for these two states can be modeled by a simple two-manifold coupled system, when neglecting contributions from other vibrational and rotational levels, from nucleus-nucleus interaction and nucleus-rotation coupling. This allows us to treat the two state manifolds separately in the following spectrally local model.\\\noindent
The excited state can be described by Hund's case (\textit{a}) and that allows to consider the total angular momentum $J=1$ as resulting from the coupling of the rotational angular momentum $\vec{N}$ and the spin $\vec{S}$ \cite{Ishikawa1993}. \Na\ and \K\ nuclear spins are $i_\textrm{\scriptsize Na}=i_\textrm{\scriptsize K}= 3/2$ leading to $3\times4\times4=48$ different states in the singlet and triplet channel.\\In absence of coupling the Hamiltonians for the investigated states, in the basis of $\ket{B/c,J,m_J,i_\mathrm{Na},m_{i_\mathrm{Na}},i_\mathrm{K},m_{i_\mathrm{K}}}$, reads as \cite{Townes1955} 

\begin{eqnarray}
H^B &= \mu_0 g_J^{B} \vec{J}\cdot \vec{B}\\\newline
H^c &= \Delta +  \vec{J} \cdot (\mathrm{A}_{\mathrm{K}} \vec{i}_\mathrm{K} + \mathrm{A}_{\mathrm{Na}} \vec{i}_\mathrm{Na}) + \mu_0 g_J^{c} \vec{J}\cdot \vec{B},
\label{Hamilt}
\end{eqnarray}\\\noindent
where $\mathrm{A}_{\mathrm{K}}$ and $\mathrm{A}_{\mathrm{Na}}$ are the hyperfine constants, $\mu_0$ the Bohr magneton and $\vec{B}$ the applied magnetic field. The energy $\Delta$ corresponds to the unperturbed energy difference between the two states and we neglect the hyperfine term for the singlet state.
The \Na\ hyperfine coupling $\mathrm{A}_\mathrm{Na}$ is almost two orders of magnitude larger than $\mathrm{A}_{\mathrm{K}}$ and it was already resolved in previous spectroscopic measurements on cold molecular beams \cite{NaKSpectroscopyTiemann2015}. The hyperfine splitting of \K\ is comparable or even smaller than the molecular state linewidth and remains mainly unresolved. At low field this allows to consider $F_1=\mid\vec{J}+\vec{i}_\textrm{\scriptsize Na}\mid=({1/2,3/2,5/2})$ as a good quantum number, as visible in Fig.\,\ref{AtomNumber}, with three groups of states at low magnetic field. Additionally, one can consider the total angular momentum $\vec{F}=\vec{J}+\vec{i}_\textrm{\scriptsize Na}+\vec{i}_\textrm{\scriptsize K}=\vec{F}_1+\vec{i}_\textrm{\scriptsize K}$, where the projection on the quantization axes $m_F$ is the only preserved quantity even at large magnetic fields.
The Zeeman term of Eq.\,\ref{Hamilt} neglects nuclear contributions and can be expanded as $\mu_0 m_J g_J B$, where $m_J$ is the projection of $\vec{J}$ on the magnetic field axis and $g_J$ the Land\'e factor. In absence of coupling between singlet and triplet, one has $g^{B}_J=g_N=g_L/(J(J+1))=1/2$ and $g^{c}_J=g_s/2\approx1$ \cite{Townes1955,SEMENOV201657}, where $g_L$ and $g_S$ are the known electron orbital and spin g-factors equal to 1 and 2.0023, respectively.
Spin-orbit interaction couples singlet and triplet with strength $\xi_{Bc}$ with selection rules $\Delta J=0$, $\Delta m_{i_\mathrm{Na}}=0$ and $\Delta m_{i_\mathrm{K}}=0$. The coupling also already incorporates the vibrational wavefunction overlap of the two states. Hence, the problem reduces to solving the following $(48+48)\times(48+48)$ matrix

\begin{eqnarray}
\eqalign{\begin{pmatrix}
    {H^{B} & \xi_{Bc}\\
    \xi_{Bc} & H^{c}}\end{pmatrix}\,.}
\end{eqnarray}\\\noindent
Note that the coupling $\xi_{Bc}$ shows its effect in two ways: 
\begin{itemize} 
	\item  As the Zeeman term remains significantly smaller than both the unperturbed singlet-triplet energy difference $\Delta$ and of the coupling strength $\xi_{Bc}$, one finds for strong fields compared to the hyperfine splitting (not shown in the later Fig.\,\ref{ATS}) the effective Land\'e factors $g^{B,c}_{J,\textrm{\scriptsize eff}}$ 
\begin{eqnarray}
g^{B,c}_{J,\textrm{\scriptsize eff}}&=\frac{1}{2} \left( g^{c}_J+g^{B}_J \mp\Delta\frac{g^{c}_J-g^{B}_J}{\sqrt{4\xi_{Bc}^2+\Delta^2}}\right)\\
&=g^{B,c}_{J} \pm \frac{1}{4}\left(  1- \frac{\Delta}{\widetilde{\Delta}}\right),
\end{eqnarray}\\\noindent
where $\widetilde{\Delta}=\sqrt{4\xi_{Bc}^2+\Delta^2}$ is the energy difference of the hyperfine centers at $B=0$ in the the coupled case.
\item The acquired triplet character of the singlet state will lead to a large hyperfine splitting, while the triplet one is decreased.
\end{itemize}

\subsection{One-photon spectroscopy}
\label{OnePhot}
\begin{figure}[ht]
\centering
\includegraphics[width=0.7\columnwidth]{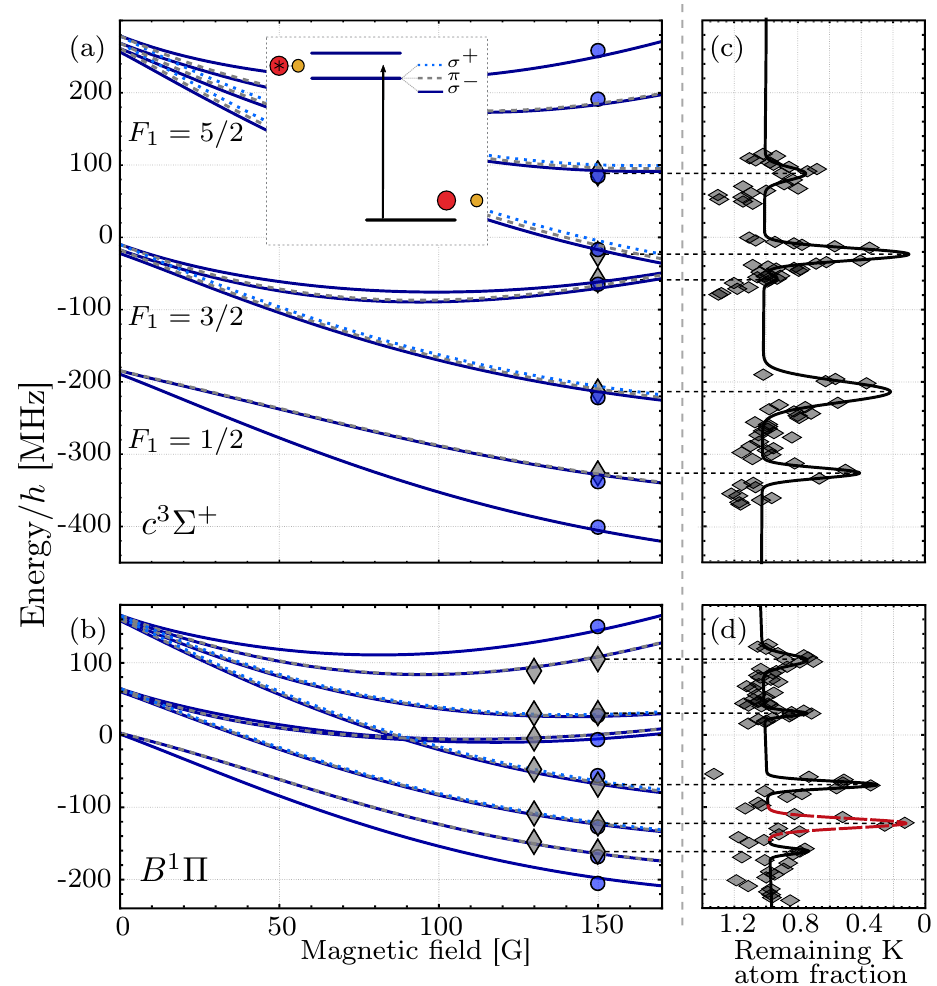}
\caption{Transitions from the atomic asymptote to the $\ket{B^1\Pi, v=8}$ and $\ket{c^3\Sigma^+, v=30}$ states. (a,b) The gray diamonds correspond to observed $\pi$-transitions and the blue circles to observed $\sigma^+$/$\sigma^-$-transitions. The lines are showing all possible transitions from the atomic states to the singlet (b) and triplet (a) states, using the described coupled model corrected by the Zeeman shift of the initial atomic states. All possible $\pi$-transitions are plotted as dashed gray lines and $\sigma^+$/$\sigma^-$-transitions as dotted light blue/solid dark blue lines. The zeros of the energy scale for the singlet and triplet transitions correspond to the hyperfine centers and are separated by the perturbed energy difference $\widetilde{\Delta}$. (c,d) Spectroscopy data for all measured $\pi$-transitions at $B=150\,\textrm{G}$. For clarity only the losses originating from $\pi$-transitions visible on \K\ are shown. A multi-Gaussian fit (solid line) is used to extract the transition energies which are used for the model fit in (a,b). (d) The red marked loss feature is used for the two-photon spectroscopy in Sec.\ref{TwoPhoton}}
\label{AtomNumber}
\end{figure}

From spectroscopic investigations \cite{NaKSpectroscopyTiemann2015} the transition energies from the atomic scattering threshold in the $a^3\Sigma^+$ potential to the dominantly $\ket{B^1\Pi, v=8}$ and dominantly $\ket{c^3\Sigma^+, v=30}$ states are expected to be about $12242.927\,\textrm{cm}^{-1}$ for the triplet state and about $12242.017\,\textrm{cm}^{-1}$ for the singlet state at a magnetic field of $150\,\textrm{G}$. In the following we will refer to the states with a dominant triplet character as \textit{triplet} and to the ones with dominantly singlet character as \textit{singlet}.\\\noindent
To locate the excited states, we measure the remaining atom number in both species after applying the Pump light for up to $1.6\,\textrm{s}$ for variable detunings from the expected transition frequencies. The atoms associated to molecules in the excited state collide with the remaining atoms and/or decay to a lower lying molecular state. Both of these effects manifest themselves as a decreased number of atoms in the initial state. The step size for the laser detuning scan is set to a few $\textrm{MHz}$, which is sufficient to resolve the different structures with an expected linewidth of about $2\pi\times6\,\textrm{MHz}$. The measurements are performed both for polarizations parallel to the magnetic field (which implies $\pi$-transitions from the atomic to the molecular state) and perpendicular to it (leading to both $\sigma^+$- and $\sigma^-$-transitions). Figure \ref{AtomNumber}(a,b) shows the positions of the loss features for the triplet and singlet states and $\pi$- and $\sigma^{+/-}$-transitions.  In the case of the singlet state, we perform spectroscopic measurements at two different magnetic field strengths to calibrate our model for the Zeeman splitting. Figure \ref{AtomNumber}(c,d) show the corresponding atom losses from $\pi$-transitions measured on \K.
\\\noindent
In the magnetic field range at which our measurements are performed, the energy splitting remains in an intermediate region between pure Zeeman and Paschen-Back regime (see Fig.\ref{AtomNumber}), where $F$ and $F_1$ are no good quantum numbers. Selection rules based on the $m_F$ allow to decrease the number of states to 18 observable ones, 6 for $\pi$- and 12 (3/9) for $\sigma^+$/$\sigma^-$-transitions. \\\noindent
Figure \ref{AtomNumber}(a,b) contains the complete collection of the observed resonant features with a distance of $27.347$\,GHz ($\mathrel{\hat=}0.913\,\textrm{cm}^{-1}$) between the manifold zeroes. The lines result from the fit of our model to the experimental observations, where the coupling $\xi_{Bc}$, the energy difference $\Delta$ and the hyperfine coupling constants $\mathrm{A}_\textrm{\scriptsize Na}$ and $\mathrm{A}_\textrm{\scriptsize K}$ are used as free parameters. The overall curvature of the lines originates from the atomic state Zeeman effect which is in the intermediate regime for both atoms. From the fit, we obtain $g^{B}_{J,\textrm{\scriptsize eff}}=0.630(8)$ and $g^{c}_{J,\textrm{\scriptsize eff}}=0.870(8)$. The uncertainty comes from the fit and from the ambiguous assignment of few resonances to a precise state, due to the simultaneous presence of $\sigma^+$/$\sigma^-$ polarized light. The unperturbed energy difference $\Delta$ is found to be equal to $13.1(8)\,\textrm{GHz}$ ($=0.437(31)\,\textrm{cm}^{-1}$) and $\xi_{Bc}$ to be equal to $12.0(2)\,\textrm{GHz}$ ($=0.400(8)\,\textrm{cm}^{-1}$). The obtained values correspond to a $26(1)\,\%/74(1)\,\%$ mixing of singlet and triplet states, which is in agreement with previous predictions \cite{Stolyarov}. Remarkably the coupling strength $\xi_{Bc}$ is comparable to the unperturbed energy difference $\Delta$. This explains the large mixing. The obtained value for $\mathrm{A}_\textrm{\scriptsize Na}$ of $307(9)\,$MHz ($\mathrel{\hat=}0.0102(3)\,\textrm{cm}^{-1}$) is in good agreement with previous spectroscopic measurements \cite{Ishikawa1992}. Our model provides a value for $\mathrm{A}_\textrm{\scriptsize K}=6(9)\,$MHz ($\mathrel{\hat=}2(3)\times10^{-3}\,\textrm{cm}^{-1}$), compatible with previous predictions and measurements \cite{Ishikawa1992, Kowalczyk1989}. \\\noindent
The amplitudes of the loss features depend on the coupling between the atomic and molecular states and are therefore highly dependent on their hyperfine quantum numbers and the wavefunction overlap. The strong mixing of the wavefunctions in the excited states and the scattering wavefunction of the atoms lead to interference effects altering the effective transition dipole strength. Our calculations do not account for the interference and thus do not provide an accurate description of transition amplitudes \cite{Schulze}.

\section{Two-photon ground state spectroscopy}
\label{TwoPhoton}
Using the located excited state as a bridge, we demonstrate two-photon spectroscopy of the rovibrational ground state $\ket{X^1\Sigma, v=0, N=0}$; as sketched in Fig.\ref{Potentials}. The energy difference between ground and excited state is well known from spectroscopic data \cite{NaKSpectroscopyTiemann2015} and can be predicted within a precision of $\pm0.08\,\textrm{cm}^{-1}$. To exactly determine the energy of the ro-vibrational ground state, we fixed the Pump laser frequency on the transition with strongest losses from Fig.\ref{AtomNumber}(d) within the singlet states which is marked with a red dashed curve. The state originates from the combination of two basis states which are labeled by $\ket{B/c,1,-1,3/2,-1/2,3/2,-1/2}$ and  $\ket{B/c,1,0,3/2,-1/2,3/2,-3/2}$ where the quantum numbers are as described in Sec.\ref{LocMod}.
\begin{figure}[ht]
	\includegraphics*[width=1\columnwidth]{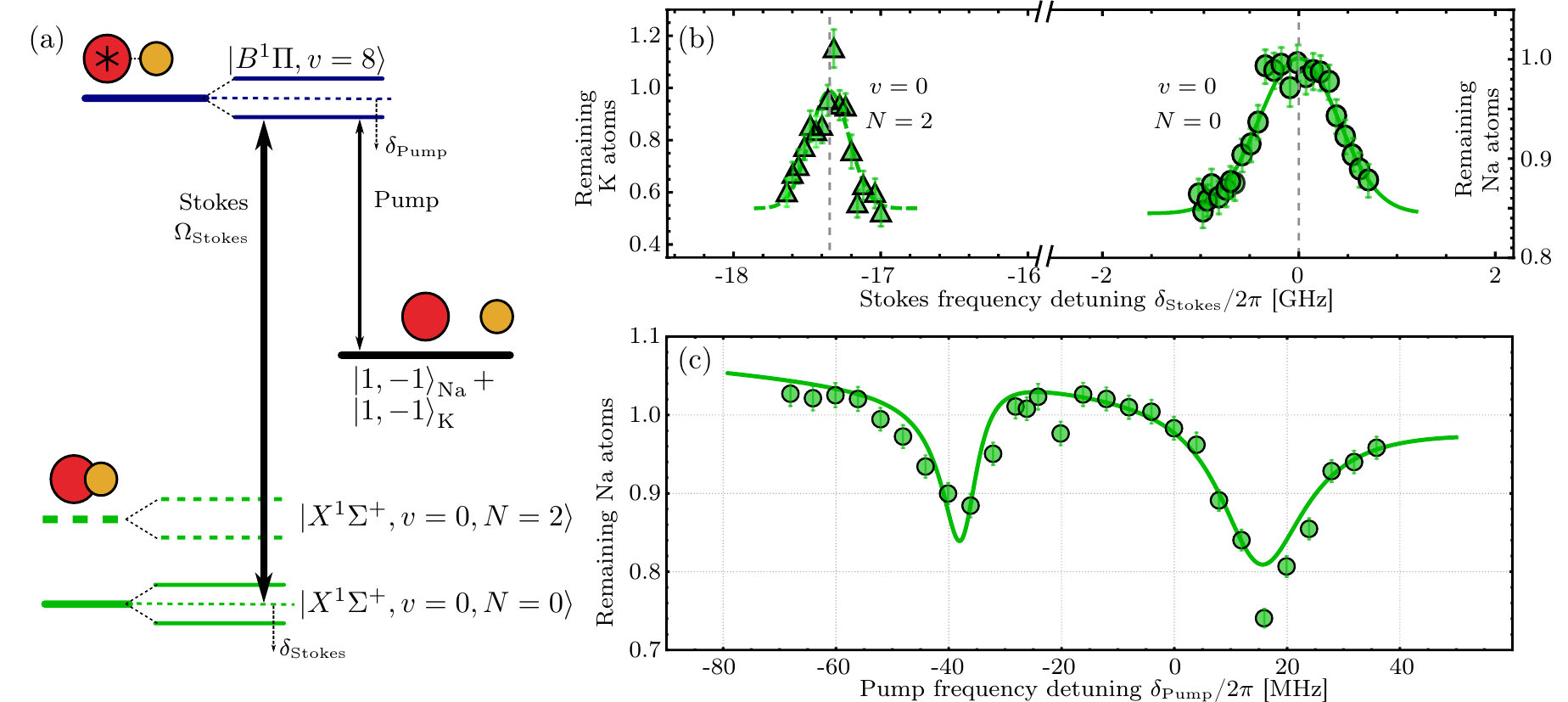}
	\caption{Two-photon spectroscopy. (a) The $\Lambda$-level scheme shows the involved energy structure for the dark-resonance spectroscopy as well as induced Stark shifts and line splitting. (b) The two lowest lying accessible rotational states of the $X^1\Sigma^+$ potential measured with dark-resonance spectroscopy. (c) Autler-Townes spectroscopy for the determination of the coupling strength between the absolute ground state and the singlet excited state. The asymmetry of the Autler-Townes structure originates from a detuning of the laser frequency compared to the precise transition energy.}
	\label{ATS}
\end{figure}\\\noindent
When the Stokes laser is on resonance, it is inducing a Stark shift on the excited state and a revival of the atom number is expected as sketched in Fig.\ref{ATS}(a). Figure \ref{ATS}(b) shows this protection when the detuning $\delta_\textrm{\scriptsize Stokes}$ is scanned with a full width half maximum for the $\ket{X^1\Sigma^+, v=0, N=0}$ of about $405(35)\,\textrm{MHz}$ for a laser power of $5\,\textrm{mW}$. The observed peak of the transition energy is at $17452.826(2)\,\textrm{cm}^{-1}$.\\\noindent
Furthermore, to unambiguously identify the ground state, we performed an atom-loss scan also in the range of frequencies where the second rotationally excited state with $N=2$ is expected; see Fig.\ref{ATS}(b). The observed energy difference of $\Delta_{\mathrm{N}=0\rightarrow\mathrm{N}=2}=h \times 17.3(3)\,\textrm{GHz}$ allows us to deduct the rotational constant to be $B_{v=0} =h\times 2.89(5)\,\textrm{GHz}$ by using the relation $\Delta_{\mathrm{N}=0\rightarrow\mathrm{N}=2}=B_0\times2(2+1)$. This value agrees with the one observed by using microwave spectroscopy \cite{MWNaK}. The full width of half maximum for the protection is $133(27)\,\textrm{MHz}$. This is smaller than for the $N=0$ state, indicating a weaker coupling to the excited states, as expected from theoretical considerations.\\\noindent 
To directly determine the coupling strength between the ground and the excited state, we fixed the Stokes laser frequency close to resonance and scanned the pump laser frequency detuning $\delta_\textrm{\scriptsize Pump}$. This scan reveals the well-known Autler-Townes splitting \cite{AutlerTownes}, the magnitude of which is proportional to the dipole matrix element of the transition. Figure \ref{ATS}(c) shows the measured remaining atom number with the typical double-loss feature. The asymmetric shape of the splitting originates from a residual detuning to the exact transition frequency of the Stokes laser which is determined to be $\delta_\textrm{\scriptsize Stokes}=2\pi\times-22\,\textrm{MHz}$. We derive the Rabi frequency $\Omega_\textrm{\scriptsize Stokes}=2\pi\times23.5\,\textrm{MHz}$ for an applied laser power of $5\,\textrm{mW}$, corresponding to a normalized Rabi frequency of $\widetilde\Omega_\textrm{\scriptsize Stokes}=2\pi\times65.2\,\textrm{kHz}\times\sqrt{I/(\textrm{mW}\,\textrm{cm}^{-2})}$ and an effective transition dipole moment of $0.170\,\textrm{D}$. Both values, $\delta_\textrm{\scriptsize Stokes}$ and $\Omega_\textrm{\scriptsize Stokes}$ are derived from a three-level master equation modeling the line shape shown in Fig.\ref{ATS}(c).
\section{Conclusion and Outlook}
\label{sec:Result}
Within this work, we have characterized a two-photon scheme for the coupling of the diatomic scattering threshold in the $a^3\Sigma^+$ potential to the rovibrational ground state $\ket{X^1\Sigma^+,v=0,N=0}$. Using photoassociation spectroscopy, we have observed and characterized the excited state hyperfine manifolds of the coupled $\ket{B^1\Pi,v=8}$ and $\ket{c^3\Sigma^+, v=30}$ states in the bosonic \NaK\ molecules at $130\ \textrm{and}\ 150\,\textrm{G}$, starting from an ultracold atomic quantum gas mixture in the states $\ket{f=1, m_f=-1}_\textrm{\scriptsize Na} + \ket{f=1, m_f=-1}_\textrm{\scriptsize K}$. By applying a spectral local model fit to the measurements, we have extracted the admixture of these states to be 26\,\%/74\,\%. Due to the strong singlet-triplet mixing, this part of the spectrum serves as an ideal bridge from the triplet atomic scattering threshold to the singlet rovibronic ground state molecules. Making use of this bridge, we have identified the rovibrational ground state and the second rotationally excited state in two-photon spectroscopy. From an Autler-Townes measurement we have extracted the Rabi-coupling between the excited and the ground state. Our work results in a fully characterized scheme for the conversion of \NaK\ Feshbach molecules to rovibrational ground state polar molecules and will allow for the efficient creation of ultracold ensembles of chemically stable bosonic \NaK\ ground state molecules.
\section*{Acknowledgments}
We thank Eberhard Tiemann for enlightening discussions and suggestions. We gratefully acknowledge financial support from the European Research Council through ERC Starting Grant POLAR and from the Deutsche Forschungsgemeinschaft (DFG) through CRC 1227 (DQ-mat), project A03 and FOR2247, project E5. K.K.V. and P.G. thank the Deutsche Forschungsgemeinschaft for financial support through Research Training Group (RTG) 1991.
\section*{References}
\bibliographystyle{iopart-num}

\begin{thebibliography}{10}
	\expandafter\ifx\csname url\endcsname\relax
	\def\url#1{{\tt #1}}\fi
	\expandafter\ifx\csname urlprefix\endcsname\relax\def\urlprefix{URL }\fi
	\providecommand{\eprint}[2][]{\url{#2}}
	
	\bibitem{IntroductionKRbChemicalReactionOspelkaus2010}
	Ospelkaus S, Ni K~K, Wang D, de~Miranda M~H~G, Neyenhuis B, Qu{\'e}m{\'e}ner G,
	Julienne P~S, Bohn J~L, Jin D~S and Ye J 2010 {\em Science\/} {\bf 327}
	853--857 ISSN 0036-8075 (\textit{Preprint}
	\eprint{http://science.sciencemag.org/content/327/5967/853.full.pdf})
	\urlprefix\url{http://science.sciencemag.org/content/327/5967/853}
	
	\bibitem{WangCollisions}
	Guo M, Ye X, He J, Gonz\'alez-Mart\'{\i}nez M~L, Vexiau R, Qu\'em\'ener G and
	Wang D 2018 {\em Phys. Rev. X\/} {\bf 8}(4) 041044
	\urlprefix\url{https://link.aps.org/doi/10.1103/PhysRevX.8.041044}
	
	\bibitem{Inelastic2010}
	Zuchowski P~S and Hutson J~M 2010 {\em Phys. Rev. A\/} {\bf 81} 060703(R)
	
	\bibitem{QuantumMagnetPolar}
	Gorshkov A~V, Manmana S~R, Chen G, Ye J, Demler E, Lukin M~D and Rey A~M 2011
	{\em Phys. Rev. Lett.\/} {\bf 107}(11) 115301
	\urlprefix\url{https://link.aps.org/doi/10.1103/PhysRevLett.107.115301}
	
	\bibitem{ThreeInteractionPolar}
	B\"uchler H, Micheli A and Zoller P 2007 {\em Nature Physics 3, 726-731\/}
	
	\bibitem{STIRAPBergmann}
	Bergmann K, Theuer H and Shore B~W 1998 {\em Rev. Mod. Phys.\/} {\bf 70}(3)
	1003--1025
	\urlprefix\url{https://link.aps.org/doi/10.1103/RevModPhys.70.1003}
	
	\bibitem{KRb1}
	Ni K~K, Ospelkaus S, de~Miranda M~H~G, Peer A, Neyenhuis B, Zirbel J~J,
	Kotochigova S, Julienne P~S, Jin D~S and Ye J 2008 {\em Science\/} {\bf 322}
	231--235
	
	\bibitem{GsDiMo23Na40K2015}
	Park J~W, Will S~A and Zwierlein M~W 2015 {\em Phys. Rev. Lett.\/} {\bf
		114}(20) 205302
	\urlprefix\url{https://link.aps.org/doi/10.1103/PhysRevLett.114.205302}
	
	\bibitem{GsDiMo23Na6Li2017}
	Rvachov T~M, Son H, Sommer A~T, Ebadi S, Park J~J, Zwierlein M~W, Ketterle W
	and Jamison A~O 2017 {\em Phys. Rev. Lett.\/} {\bf 119}(14) 143001
	\urlprefix\url{https://link.aps.org/doi/10.1103/PhysRevLett.119.143001}
	
	\bibitem{GsDiMo87Rb133Cs2014Grimm}
	Takekoshi T, Reichs\"ollner L, Schindewolf A, Hutson J~M, Le~Sueur C~R, Dulieu
	O, Ferlaino F, Grimm R and N\"agerl H~C 2014 {\em Phys. Rev. Lett.\/} {\bf
		113}(20) 205301
	\urlprefix\url{https://link.aps.org/doi/10.1103/PhysRevLett.113.205301}
	
	\bibitem{GsDiMo87Rb133Cs2014Cornish}
	Molony P~K, Gregory P~D, Ji Z, Lu B, K\"oppinger M~P, Le~Sueur C~R, Blackley
	C~L, Hutson J~M and Cornish S~L 2014 {\em Phys. Rev. Lett.\/} {\bf 113}(25)
	255301
	\urlprefix\url{https://link.aps.org/doi/10.1103/PhysRevLett.113.255301}
	
	\bibitem{NaRb1}
	Guo M, Zhu B, Lu B, Ye X, Wang F, Vexiau R, Bouloufa-Maafa N, Qu\'em\'ener G,
	Dulieu O and Wang D 2016 {\em Phys. Rev. Lett.\/} {\bf 116}(20) 205303
	\urlprefix\url{https://link.aps.org/doi/10.1103/PhysRevLett.116.205303}
	
	\bibitem{Frauke}
	See\ss{}elberg F, Buchheim N, Lu Z~K, Schneider T, Luo X~Y, Tiemann E, Bloch I
	and Gohle C 2018 {\em Phys. Rev. A\/} {\bf 97}(1) 013405
	\urlprefix\url{https://link.aps.org/doi/10.1103/PhysRevA.97.013405}
	
	\bibitem{NaKSpectroscopyTiemann2015}
	Temelkov I, Kn\"ockel H, Pashov A and Tiemann E 2015 {\em Phys. Rev. A\/} {\bf
		91}(3) 032512
	\urlprefix\url{https://link.aps.org/doi/10.1103/PhysRevA.91.032512}
	
	\bibitem{Schulze}
	Schulze T~A, Temelkov I~I, Gempel M~W, Hartmann T, Kn\"ockel H, Ospelkaus S and
	Tiemann E 2013 {\em Phys. Rev. A\/} {\bf 88}(2) 023401
	\urlprefix\url{https://link.aps.org/doi/10.1103/PhysRevA.88.023401}
	
	\bibitem{Aymar2017}
	Aymar M and Dulieu O 2007 {\em Molecular Physics\/} {\bf 105} 1733--1742
	(\textit{Preprint} \eprint{https://doi.org/10.1080/00268970701494016})
	\urlprefix\url{https://doi.org/10.1080/00268970701494016}
	
	\bibitem{AutlerTownes}
	Autler S~H and Townes C~H 1955 {\em Phys. Rev.\/} {\bf 100}(2) 703--722
	\urlprefix\url{https://link.aps.org/doi/10.1103/PhysRev.100.703}
	
	\bibitem{Gempel}
	Gempel M~W 2016 Ph.D. thesis University of Hannover
	
	\bibitem{Hartmannphd}
	Hartmann T 2018 Ph.D. thesis University of Hannover
	
	\bibitem{SchulzeBEC2018}
	Schulze T~A, Hartmann T, Voges K~K, Gempel M~W, Tiemann E, Zenesini A and
	Ospelkaus S 2018 {\em Phys. Rev. A\/} {\bf 97}(2) 023623
	\urlprefix\url{https://link.aps.org/doi/10.1103/PhysRevA.97.023623}
	
	\bibitem{Hartmann1}
	Hartmann T, Schulze T~A, Voges K~K, Gersema P, Gempel M~W, Tiemann E, Zenesini
	A and Ospelkaus S 2019 {\em Phys. Rev. A\/} {\bf 99}(3) 032711
	\urlprefix\url{https://link.aps.org/doi/10.1103/PhysRevA.99.032711}
	
	\bibitem{Drever1983}
	Drever R~W~P, Hall J~L, Kowalski F~V, Hough J, Ford G~M, Munley A~J and Ward H
	1983 {\em Applied Physics B\/} {\bf 31} 97--105 ISSN 1432-0649
	\urlprefix\url{https://doi.org/10.1007/BF00702605}
	
	\bibitem{Gregory_2015}
	Gregory P~D, Molony P~K, Köppinger M~P, Kumar A, Ji Z, Lu B, Marchant A~L and
	Cornish S~L 2015 {\em New Journal of Physics\/} {\bf 17} 055006
	\urlprefix\url{https://iopscience.iop.org/article/10.1088/1367-2630/17/5/055006}
	
	\bibitem{Ishikawa1993}
	Ishikawa K 1993 {\em The Journal of Chemical Physics\/} {\bf 98} 1916--1924
	
	\bibitem{Townes1955}
	Townes C~H and Schawlow A~L 1955 {\em Microwave spectroscopy\/} (New York:
	Dover)
	
	\bibitem{SEMENOV201657}
	Semenov M, Yurchenko S~N and Tennyson J 2016 {\em Journal of Molecular
		Spectroscopy\/} {\bf 330} 57--62 ISSN 0022-2852 potentiology and Spectroscopy
	in Honor of Robert Le Roy
	
	\bibitem{Stolyarov}
	Ferber R, Pazyuk E~A, Stolyarov A~V, Zaitsevskii A, Kowalczyk P, Chen H, Wang H
	and Stwalley W~C 2000 {\em J. Chem. Phys.\/} {\bf 112} 5740
	
	\bibitem{Ishikawa1992}
	Ishikawa K, Kumauchi T, Baba M and Kat\^o H 1992 {\em The Journal of Chemical
		Physics\/} {\bf 96} 6423--6432 (\textit{Preprint}
	\eprint{https://doi.org/10.1063/1.462856})
	\urlprefix\url{https://doi.org/10.1063/1.462856}
	
	\bibitem{Kowalczyk1989}
	Kowalczyk P 1989 {\em The Journal of Chemical Physics\/} {\bf 91} 2779--2789
	
	\bibitem{MWNaK}
	Yamada C and Hirota E 1992 {\em Journal of Molecular Spectroscopy\/} {\bf 153}
	91
	
\end{thebibliography}
\providecommand{\newblock}{}

\end{document}